\documentclass[superscriptaddress,twocolumn,pra,10pt,longbibliography]{revtex4-1} 
\usepackage{graphicx}
\usepackage{epsfig}
\usepackage[caption=false]{subfig}
\usepackage{verbatim}
\usepackage{amssymb}
\usepackage{amsmath}
\usepackage{amsfonts}
\usepackage{refcount}
\usepackage{bm}
\usepackage{hyperref}
\usepackage{color}
\usepackage[margin=.7in,nofoot]{geometry}
\usepackage{hyperref}
\hypersetup{
    colorlinks=true,       
    linkcolor=cyan,          
    citecolor=magenta,        
    filecolor=magenta,      
    urlcolor=cyan,           
    runcolor=cyan
}
\usepackage{algorithm}
\usepackage{algpseudocode}

\newcommand {\be}{\begin{equation}}
\newcommand {\ee}{\end{equation}}
\newcommand {\bea}{\begin{eqnarray}}
\newcommand {\eea}{\end{eqnarray}}

\begin{document}

\title{Practical engineering of hard spin-glass instances}

\author{Jeffrey Marshall}
\affiliation{Department of Physics and Astronomy, University of Southern California, Los Angeles,
CA 90089, USA}
\affiliation{Center for Quantum Information Science \& Technology, University of Southern California, Los Angeles, CA 90089, USA}
\email{jsmarsha@usc.edu}
\author{Victor Martin-Mayor}
\affiliation{Departamento de F\'isica Te\'orica I, Universidad Complutense, 28040 Madrid, Spain}
\affiliation{Instituto de Biocomputaci\'on y F\'isica de Sistemas Complejos (BIFI), Zaragoza, Spain}
\author{Itay Hen}
\affiliation{Center for Quantum Information Science \& Technology, University of Southern California, Los Angeles, CA 90089, USA}
\affiliation{Information Sciences Institute, University of Southern California, Marina del Rey, CA 90292, USA}
\begin{abstract}
Recent technological developments in the field of experimental quantum annealing have made prototypical annealing optimizers with hundreds of qubits commercially available. The experimental demonstration of a quantum speedup for optimization problems has since then become a coveted, albeit elusive goal. Recent studies have shown that the so far inconclusive results, regarding a quantum enhancement, may have been partly due to the benchmark problems used being unsuitable. In particular, these problems had inherently too simple a structure, allowing for both traditional resources and quantum annealers to solve them with no special efforts. The need therefore has arisen for the generation of harder benchmarks which would hopefully possess the discriminative power to separate classical scaling of performance with size, from quantum. We introduce here a practical technique for the engineering of extremely hard spin glass Ising-type problem instances that does not require `cherry picking' from large ensembles of randomly generated instances. We accomplish this by treating the generation of hard optimization problems itself as an optimization problem, for which we offer a heuristic algorithm that solves it. We demonstrate the genuine thermal hardness of our generated instances by examining them thermodynamically and analyzing their energy landscapes, as well as by testing the performance of various state-of-the art algorithms on them. We argue that a proper characterization of the generated instances offers a practical, efficient way  to properly benchmark experimental quantum annealers, as well as any other optimization algorithm.  
\end{abstract}

\maketitle

\section{Introduction}
Many problems of theoretical and practical relevance consist of searching for the global minimum of a cost function. These optimization problems are on the one hand notoriously hard to solve but on the other hand ubiquitous, and appear in diverse fields such as machine learning, material design, and software verification, to mention a few diverse examples. The computational difficulties associated with solving these optimization problems stems from the intricate structure of the cost function that needs to be optimized which often has a rough landscape with many local minima. The design of fast and practical algorithms for optimization has therefore become one of the most important challenges of many areas of science and technology~\cite{papadimitriou2013combinatorial}.
\par
Recent theoretical and technological breakthroughs have triggered an enormous interest in one such non-traditional method, commonly referred to as Quantum Annealing (QA)~\cite{finnila_quantum_1994,kadowaki_quantum_1998,sqa1,Santoro,Brooke1999,farhi_quantum_2001,Reichardt:2004}.  The uniqueness of this approach stems from the fact that it does not rely on traditional computation resources but rather manipulates data structures called quantum bits, or qubits, that obey the laws of Quantum Mechanics. 
It is believed that by utilizing uniquely quantum features such as entanglement, massive parallelism and tunneling, a quantum computer can solve certain computational problems in a way which scales better with problem size than is possible on a classical machine. 
\par
A huge amount of progress has recently been made in the building of experimental quantum annealers~\cite{DWave,Bunyk:2014hb}, the most notable of which are the D-Wave processors consisting of hundreds of coupled superconducting flux qubits. These devices offer a very natural approach to solving optimization problems utilizing gradually decreasing quantum fluctuations to traverse the barriers in the energy landscape in search of global optima~\cite{finnila_quantum_1994,Brooke1999,kadowaki_quantum_1998,farhi_quantum_2001,Santoro}. 
As an inherently quantum technique, QA holds the so-far unfulfilled promise to solve combinatorial optimization problems faster than traditional `classical' 
algorithms~\cite{young2008,hen:11,hen:12,farhi:12}. However, to date there is no experimental  (nor theoretical) evidence that quantum annealers are capable of producing such speedups~\cite{speedup,hen:15}. 
\par
Extensive studies designed to properly benchmark experimental QA processors, such as the D-Wave annealers, have resulted for the most part in inconclusive results, despite accumulating evidence for the (indirect detection) of genuinely quantum effects such as entanglement and multi-qubit tunneling~\cite{DWave-entanglement,Boixo:2014yu,q108,googleTunneling,googleTunnelingII}. Indeed, direct comparison tests between the $128$-qubit D-Wave One and later the $512$-qubit D-Wave Two (DW2) processors and classical state of the art algorithms on randomly-generated Ising-model instances have shown no evidence of a quantum speedup~\cite{q108,speedup,hen:15}.
\par
The above lack of evidence has motivated a few recent studies to further explore problem classes where one might expect the occurrence of quantum speedups~\cite{hen:15,2014Katzgraber,scirep15:Martin-Mayor_Hen,Venturelli:2014nx}. 
Katzgraber, Hamze and Andrist~\cite{2014Katzgraber} pointed out that the
random Ising instances used in the previously mentioned comparison tests
exhibit a spin-glass phase transition only at $T=0$, i.e., at zero temperature. Spin-glasses are disordered, frustrated
spin systems that may be viewed as prototypical classically-hard (also called
NP-hard) optimization problems, that are so challenging that specialized
hardware has been built to simulate them~\cite{janus:08,janus:09,janus:14}
[the related cost function is in Eq.~\eqref{Ising} below]. That the spin glass transition occurs at $T=0$ implies that for any $T>0$, the energy landscapes for these problems are in general fairly simple
and can therefore be solved rather easily
by classical heuristic solvers and hence do not require quantum tunneling reach global optima, thus rendering these instances less than ideal for benchmarking. 

\par
A subsequent study which examined the same class of uniformly-random Ising problems on the D-Wave architecture~\cite{scirep15:Martin-Mayor_Hen} measured the correlation between the performance of the DW2 device and a physical effect
referred to as \emph{temperature
  chaos}~\cite{mckay:82,bray:87b,banavar:87,kondor:89,kondor:93,billoire:00,rizzo:01,mulet:01,billoire:02,krzakala:02,rizzo:03,parisi:10,sasaki:05,katzgraber:07,fernandez:13,billoire:14},
which has recently been identified as the culprit for the difficulties that classical thermal algorithms encounter when attempting to solve
spin-glasses~\cite{Fischer,Binder86}.
Temperature chaos implies the
presence of low-lying excited states (i.e., slightly sub-optimal spin
assignments) that have a large Hamming distance with respect to the minimizing assignment, or ground state (GS), of the instance. Furthermore, these excited states are not only stable against local
excitations (i.e., bit flips), they also have a much larger entropy than the
GS. As a consequence, classical state-of-the-art optimization algorithms such as simulated annealing or parallel tempering
simulations often get trapped in one of these excited states. Indeed, non-chaotic problem instances are exponentially unlikely to be found as the problem size is increased \cite{parisi:10,fernandez:13,billoire:14}.
However, while temperature-chaotic instances do indeed exist on the relatively tiny DW2 $512$-bit hardware graph, they become exceedingly rare with the degree to which they exhibit temperature chaos, and are therefore difficult to find~\cite{scirep15:Martin-Mayor_Hen}.

Since these are the temperature-chaotic instances that are expected to have the discriminative power to separate classical scaling of performance with size, from quantum, a natural question thus arises. Can one efficiently find or generate `rare gem' instances, i.e., small size
problems (so small that encoding it on a quantum device is feasible) that also
display a large degree of temperature chaos, or inherent hardness? 
To date, several techniques for generating hard problems that go beyond random generation of instances have been explored, such as utilizing instances with planted solutions with tunable frustration~\cite{hen:15}, the deliberate reduction of GS degeneracy using Sidon sets for the couplings~\cite{katzgraber:seekingSpeedup} or the porting of fully-connected Sherrington-Kirkpatrick (SK) instances~\cite{Venturelli:2014nx}. However the obtained instances were found to lack the necessary degree of inherent hardness (i.e., hard problems are still rare), which 
as a result necessitated the generation of
an initial huge pool of problems followed by the prohibitively expensive procedure of exhaustive `mining' for instances presenting high degrees of inherent thermal hardness~\cite{scirep15:Martin-Mayor_Hen,katzgraber:seekingSpeedup}.
\par
In this paper, we propose an altogether different, adaptive algorithm to generate such rare gem instances ---  extremely hard spin-glass instances of relatively small size ---  in a considerably more efficient manner than current mining techniques.
The remainder of the work is organized as follows. In Sec.~\ref{sect:spinGlass} we present the basics of a heuristic algorithm which generates hard spin-glass instances. Following this, in Sec.~\ref{sect:results}, we thermodynamically analyze the  generated instances, and test them against classical optimizers as well the DW2 annealer. We also apply our technique to instances with planted solutions in Sec.~\ref{sect:planted} and conclude in Sec.~\ref{sect:conc}.

\section{Engineering of hard spin-glass benchmarks}
\label{sect:spinGlass}
For concreteness in what follows we apply our technique to the D-Wave Two `Chimera' hardware graph (see Appendix \ref{app:dwave}) as this will allow us to experimentally test the hardness of the instances on an actual quantum annealer. 
However it should be noted that the method proposed here is far more general and may apply to arbitrary connectivity graphs. The cost function on which we generate our instances is of the form
\begin{equation}
H = \sum_{\langle i,j\rangle}J_{ij}s_i s_j,
\label{Ising}
\end{equation}
where the couplings $\{ J_{ij} \}$ are programmable parameters that define the instance. The cost function $H$
is to be minimized over the spin variables, $s_i =\pm 1$ where $i=1\ldots N$ and $N$ is the number of participating spins. The angle brackets $\langle i,j\rangle$ denote that the sum is only over connected bits on the Chimera graph.

In this work, we shall treat the process of finding problem instances, i.e., sets of $\{ J_{ij}\}$ values, over any predetermined set of allowed values in a way that maximizes the hardness of the problem (however hardness is defined), as an optimization problem in itself.
The figure of merit -- or cost function -- for this optimization problem is any faithful characteristic of the inherent hardness of the instance. We will call this figure of merit the \emph{time to solution}, or TTS for short.
\par
Here, we shall use as the TTS of a problem instance the definition for classical thermal hardness that was introduced in Ref.~\cite{scirep15:Martin-Mayor_Hen}, namely, the characteristic number of steps it takes for a parallel tempering (PT) algorithm to equilibrate. In a PT algorithm, one simulates $N_T$ realizations of an $N$-spin system, with temperatures $T_1<T_2<\ldots< T_{N_T}$, where Metropolis updates occur independently for each copy. Each copy attempts to swap temperatures with its temperature neighbors, with probabilities satisfying detailed balance~\cite{sokal:97}. The resulting temperature random walk of each system copy allows a global traversal of the configuration space, as well as detailed exploration of local minima (i.e., at the lower temperatures). An accurate  PT simulation takes time longer than the temperature `mixing' time, $\tau$ ~\cite{fernandez:09b,janus:10}. The time $\tau$ can be thought of as an equilibration time; the time for each copy to explore the entire temperature mesh. Thus, large $\tau$ instances take longer to equilibrate, motivating the definition
of the mixing time $\tau$ as the \emph{classical hardness}.

Furthermore, we shall utilize the strong correlation found between the PT mixing time $\tau$ and the hardness of other algorithms (consistently with the requirement of intrinsic hardness~\cite{scirep15:Martin-Mayor_Hen}). Specifically, we shall use as TTS the runtime clocked by the Hamze-de Freitas and Selby algorithm (HFS)~\cite{hamze:04,Selby:2014tx},  which has proved to be much faster yet strongly correlated with the classical hardness, $\tau$~\cite{scirep15:Martin-Mayor_Hen}. 
Our aim in this work is the \emph{maximization} of the TTS cost function where the variables over which the maximization is done are the coupling strengths $\{J_{ij}\}$ of the underlying graph.

\subsection{Random adaptive optimization (RAO)}
Heuristic optimization algorithms, such as Metropolis and simulated annealing, aim to find the global minimum (equivalently, maximum) of a cost function by changing the state of the system at each step. Changes are accepted when the cost moves in the required direction, but also, often crucially, still accept changes in the `wrong' direction with a certain probability so as to reduce the chances of becoming stuck in local minima. This acceptance probability may be determined by defining a simulated `temperature' parameter, $\beta$ (inspired by thermal annealing). We follow this approach with the cost being the TTS (which is assumed to be an indicator of inherent hardness), and the `state' of the system a particular configuration of the $J_{ij}$.
\par
By picking some random `seed' instance, and modifying a subset of the $J_{ij}$ (e.g., flipping the sign of a random edge), the resulting instance may be harder to solve as determined by the TTS cost. If this is the case, we accept the modification. Repeating this process will necessarily drive the system to harder and harder instances. If the TTS is lowered by such a modification, it may still be accepted, so as avoid getting trapped in local minima. We utilize a Boltzmann-type acceptance probability, $e^{-\beta |\Delta \text{TTS}|}$~\footnote{The `Boltzmann factor' $e^{-\beta |\Delta \text{TTS}|}$ should not be confused with the standard one in Statistical Mechanics, $e^{-\beta H}$: in our case the TTS plays the role that the Hamiltonian would play in a standard problem.}, where the choice of $\beta$ defining this distribution will depend on the solver being used to determine the TTS, and the manner in which one updates $\beta$ during the algorithm (if at all), will depend on the methodology one wishes to pursue (e.g., Metropolis, simulated annealing, etc.). 
\par
We outline our algorithm in its most basic (unoptimized) form in Algorithm~\ref{RAO}, which generates hard signed (i.e., $J_{i,j}=\pm 1$) instances, though we wish to stress that the general  technique can be applied under much more diverse settings. In fact, we expect that allowing the coupling constants, $J_{ij}$, to take on a wider range of values, will in general result in harder instances being generated (compared to the $J_{ij}=\pm 1$ case). However, this is not necessarily indicative of a greater efficiency of the RAO algorithm, as higher range (random) instances are known to be harder to solve, compared to $J_{ij}=\pm 1$ instances~\cite{speedup}.
Additionally, as an example of this diversity, we have also utilized a `reversed' version of the algorithm, tweaked to \emph{minimize} the TTS in order to generate particularly easy instances as well. Moreover, we apply this method to instances with planted solutions in Sec.~\ref{sect:planted}.
In the next section we illustrate the effectiveness of our technique.

\begin{algorithm}[H]
\caption{Random Adaptive Optimization (RAO)}
\label{RAO}
\begin{algorithmic}[1]
\Procedure{GenerateHardProblem}{}
\State Generate random $\pm 1$ seed instance and get TTS
\For{step = 1 \textbf{to} NSTEP }
\State Pick a random edge and flip sign
\State Get new TTS
\If{TTS increases}
\State Accept Change
\Else
\State Accept with probability $e^{-\beta |\Delta \text{TTS}|}$
\EndIf
\State Update $\beta$ if required
\EndFor
\EndProcedure
\end{algorithmic}
\end{algorithm}

\section{Results}
\label{sect:results}
\subsection{Engineering of hard spin-glass instances}

For our work, we used extensively the version of the HFS algorithm created by A. Selby~\cite{selby:13a,Selby:2014tx}, and have taken the average wallclock time for the TTS~\cite{tts_hfs}, as our cost function. We adopt the notation $t_{HFS}$ for this quantity (see Appendix \ref{app:implementation} for specific implementation details)~\footnote{If one were to use a solver which suffers from intrinsic control errors (e.g., an analogue device, such as the DW annealers), i.e., encoding errors in the $J_{ij}$, one may have to perform some kind of averaging procedure to try to estimate the TTS more accurately (e.g., in the DW case, running over several different gauges \cite{speedup}). The performance of our algorithm will of course be adversely affected in such a case.}.
\par
\begin{figure}[h]
\centering
\includegraphics[scale=0.47]{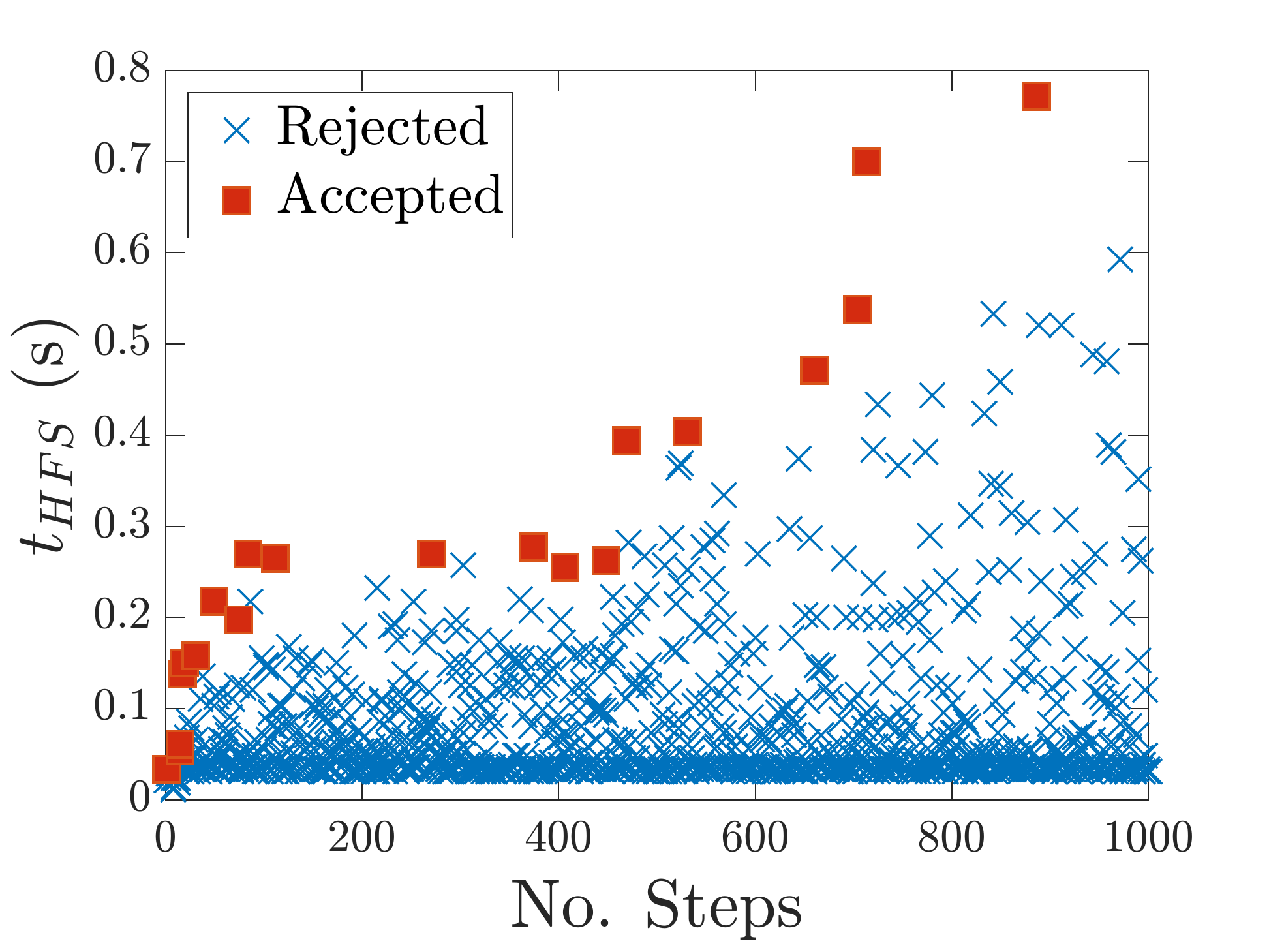}
\caption{(Color online) {\bf Performance of the RAO algorithm for a single instance}. A single run of the algorithm over 1000 steps, set up to maximize $t_{HFS}$ of a $512$-bit Chimera instance with $J_{ij}=\pm 1$. Updates consist of flipping the value of randomly picked edges. Squares (red) show successful update attempts, crosses (blue) are rejected updates.}
\label{TTS:graph}
\end{figure}

The performance of one typical realization of our algorithm on a 512 bit Chimera-type instance is depicted in Fig.~\ref{TTS:graph}. Remarkably, the final instance is just 20 successful steps away from the initial, and the final TTS is about 25 times the initial instances TTS. Though there does not seem to necessarily be a typical (or standard) output for the algorithm, the occurrence of plateaus is found to be fairly common. Indeed in Fig.~\ref{TTS:graph} we see that the instance remains at a plateau of about 0.25$s$ between 100 and 400 attempted flips.
These plateaus can occasionally halt the optimization of the cost function, and as such it is important to carefully choose an appropriate simulated temperature.

\par
\begin{figure}[h]
\centering
\includegraphics[scale=0.47]{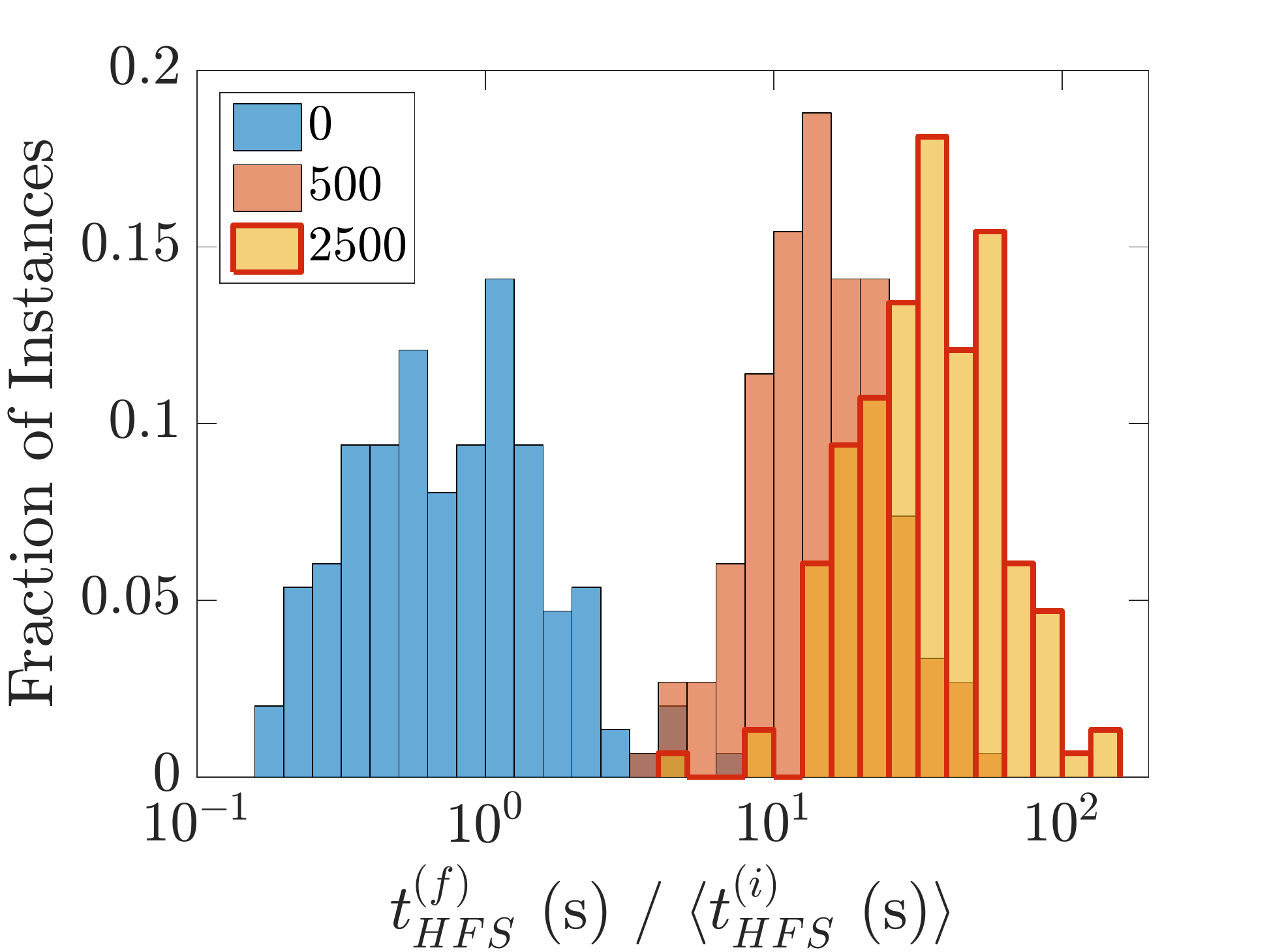}
\caption{(Color online) {\bf Histogram of the ratio of final to mean initial
    $t_{HFS}$ for 150 random signed instances (512 bit Chimera graph)}. As
  in Fig \ref{TTS:graph} we adapted the instances by flipping random edge
  values, attempting to maximize $t_{HFS}$. This plot is a normalized
  histogram of the ratio final TTS, $t_{HFS}^{(f)}$, to mean initial TTS,
  $\langle t_{HFS}^{(i)}\rangle$, after 0 (blue), 500 (red) and 2500 (yellow, with red outline) algorithmic steps.}
\label{histogram:graph}
\end{figure}

In Fig.~\ref{histogram:graph}, we statistically quantify the merit of the generated hard instances, by comparing the final TTS to the mean initial TTS (i.e., typical TTS of a random instance), after 0 (blue), 500 (red) and 2500 (yellow, with red outline) adaptive steps on 150 instances. One notices immediately a clear separation in hardness classification from the completely random instances (blue), and the other two groups, even after a fairly modest number of update attempts (i.e., 500). The instances after 2500 steps are on average about 3 times harder than those after 500 steps, which are themselves about an order of magnitude above the random instances.

\par
To gain a reverse effect, namely, easier than random instances, we have also run our algorithm by `reversing' the acceptance criterion (i.e., step 6 in Algorithm~\ref{RAO}) such that it favors instances with shorter TTS values. This allowed for the lowering the TTS on the DW2 Chimera by a factor of about 10 from randomly chosen instances (see next subsection).

\subsection{Inherent (thermal) hardness of the generated instances}\label{subsect:PT}

It is crucially important to demonstrate that the generated instances are not only difficult to solve with respect to the solver with the help of which they are obtained, but that the instances are \emph{inherently} difficult, i.e., they possess inherent degrees of hardness. In what follows we illustrate precisely that by measuring the thermodynamical complexity of the instances, generated using our RAO algorithm.

To that aim, we have used $t_{HFS}$ as both a minimizing and maximizing cost function, to generate 100 signed ($J_{ij}=\pm1$) $500$-bit DW2 Chimera instances in each of four different hardness groups, or generations, classified by \hbox{$t_{HFS}\in$ [0.8,1.2]$\times 10^{-4+k}\,(s)$}, where $k=1,2,3,4$~\cite{RAO_generation}. The $k=1$ instances are about an order of magnitude easier compared to random instances. The hardest instance we have found on the studied graph, after analyzing $780$ instances, was found to be $\sim 250$ times harder than a typical random instance, with a runtime of $t_{HFS}\approx 6.0s$ on a 3.5 GHz single core CPU, which to our knowledge is the most difficult randomly generated HFS instance on the DW2 graph to date. We shall denote this instance by $k_{max}$.

Our first task is assessing how difficult it is for a state-of-the-art classical thermal algorithms such as PT
to solve the generated instances. This question can be quantitatively answered by
computing the mixing time, $\tau$, for the
temperature random-walk of the algorithm~\cite{sokal:97,fernandez:09b,janus:10,fernandez:13} and examine its correlation with `hardness group' $k$.  

Our computation of $\tau$ follows exactly the procedure detailed
in Ref.~\cite{scirep15:Martin-Mayor_Hen}. We simulated 120 system copies consisting of four independent parallel-temperature chains,
with 30 temperatures each. Given the similar system size, we also use the
same temperature grid of Ref.~\cite{scirep15:Martin-Mayor_Hen}. The Elementary
Monte Carlo Step (EMCS) consisted of 10 full-lattice Metropolis sweeps,
independently performed in each system copy, followed by one Parallel
Tempering temperature-exchange sweep.  A computation of $\tau$ was considered
satisfactory if two conditions were met: i) The system copy (out of the 120
possibilities) that spends the least time in the hot-half region (i.e., the 15
highest temperatures), spends at least 20\% of the total simulation time
there. In other words, no system-copy got permanently trapped in the cold-half
region. ii) The total simulation time was at least $20 \tau$ long.
$\tau$ is given in units of Metropolis
sweeps.

We performed three independent simulations of different lengths:
$10^6$ EMCS (i.e., $10^7$ Metropolis sweeps), $10^7$ EMCS and $10^8$
EMCS. The shortest runs ($10^6$ EMCS), were enough to compute $\tau$
for the 200 problem instances that belong to the first and second
hardness groups ($k=1,2$). The $10^7$ EMCS run sufficed to compute $\tau$
for most (but not all) of the third-generation instances. The $10^8$
EMCS run was enough to compute $\tau$ for all the third-generation
instances, and for 86 out of 100 problem instances belonging to the
fourth-generation. As a cross-check, we compared the GS
energy found with parallel tempering with the one found with the HFS code. 
Agreement was reached in all cases (even in cases where the computation of
$\tau$ was not successful).
\par
Specifically notable are 32 of the instances of the hardest ($k=4$) HFS group, which were found to have $\tau > 10^7$. In Ref.~\cite{scirep15:Martin-Mayor_Hen} it was estimated that only 2 out of every $10^4$ random instances of this type have $\tau>10^7$. This observation helps to quantify the efficiency of our algorithm; a highly optimized PT algorithm screening random instances of this type would require $\sim 65$ CPU hours~\cite{PT_details} to find one with $\tau >10^7$. Equivalently for (unoptimized) RAO, we estimate less than $5$ hours to generate one such instance~\cite{RAO_generation}.

\begin{figure}[htp]
\centering
\includegraphics[scale=0.47]{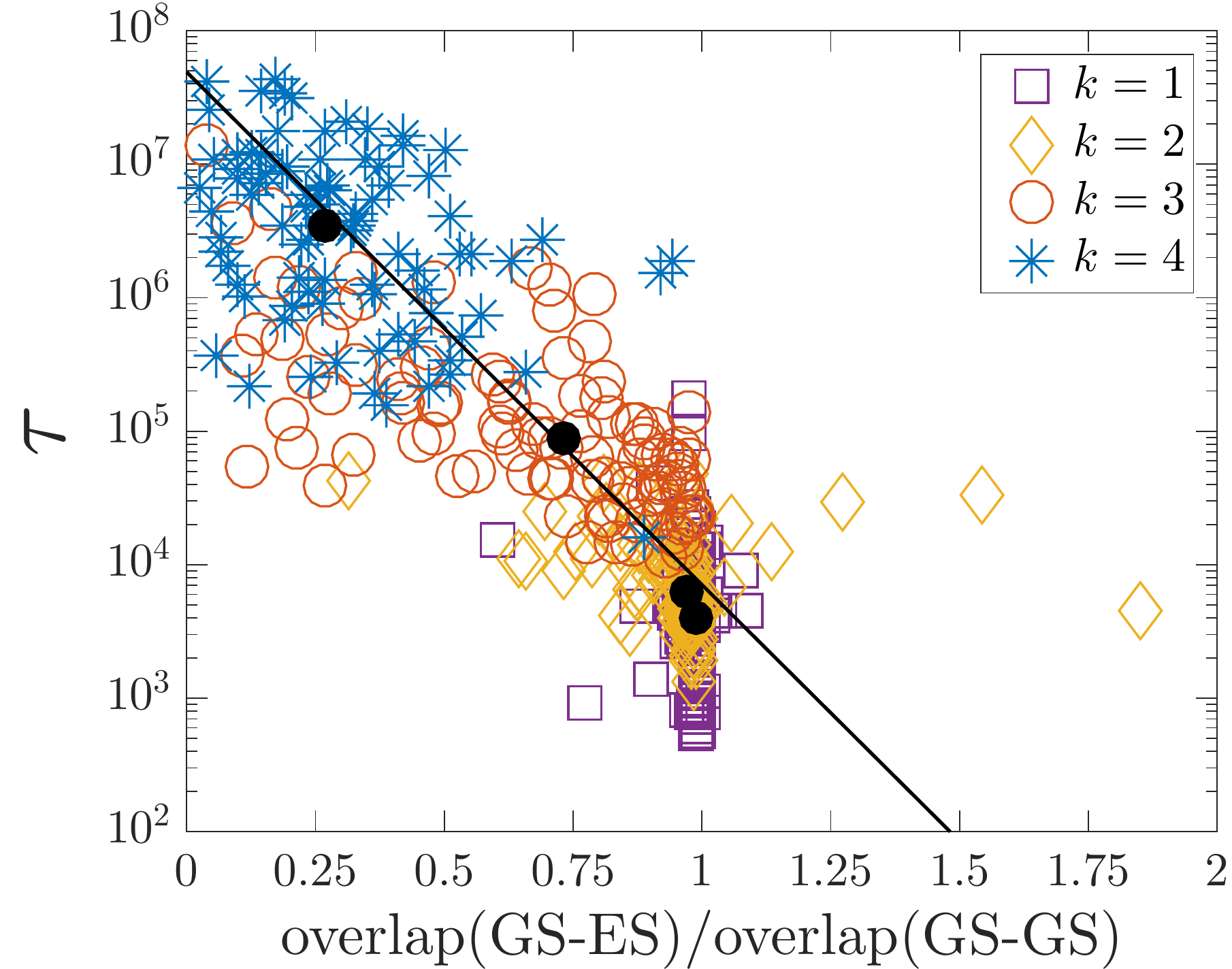}
\caption{(Color online) {\bf Mixing time $\tau$ vs ratio of typical GS-ES overlaps over GS-GS overlaps}. For each instance for which PT computed $\tau$ successfully, we compute the median overlap (see text) between the ground and dominant first-excited states, normalizing by the median ground state to ground state overlap. Also plotted is the median data point for each group (filled black circle), where going from bottom right to upper left, is in order from $k=1$ to $k=4$. Linear fit is on the median data point for $k=2,3,4$. $\tau$ is given in units of Metropolis sweeps.}
\label{overlap:graph}
\end{figure}
\par

Having classified (most of) these 400 instances by the PT mixing time, we analyze their energy landscapes by computing the overlap between their GS configurations and their dominant first-excited states. The overlap between two spin assignments, $\bar s_1,\bar s_2$, is defined as, $1-2h(\bar s_1,\bar s_2)/N$, where $h$ is the Hamming distance. This analysis is summarized in Fig.~\ref{overlap:graph}. The trend in the figure is rather clear: a large $\tau$ (and $k$) correlates strongly with a smaller overlap (i.e., large Hamming distance) between the ground and dominant first-excited states; that is, the larger $\tau$, the more difficult the problems are in a \emph{thermodynamical} sense. Interestingly, the easiest HFS instances, $k=1$, which have been generated by minimization of TTS have been found to not correlate as well with the other data groups. We examine this in more detail in the next subsection.

\subsection{Algorithmic scaling}
\begin{figure}[h]
\centering
\includegraphics[scale=0.47]{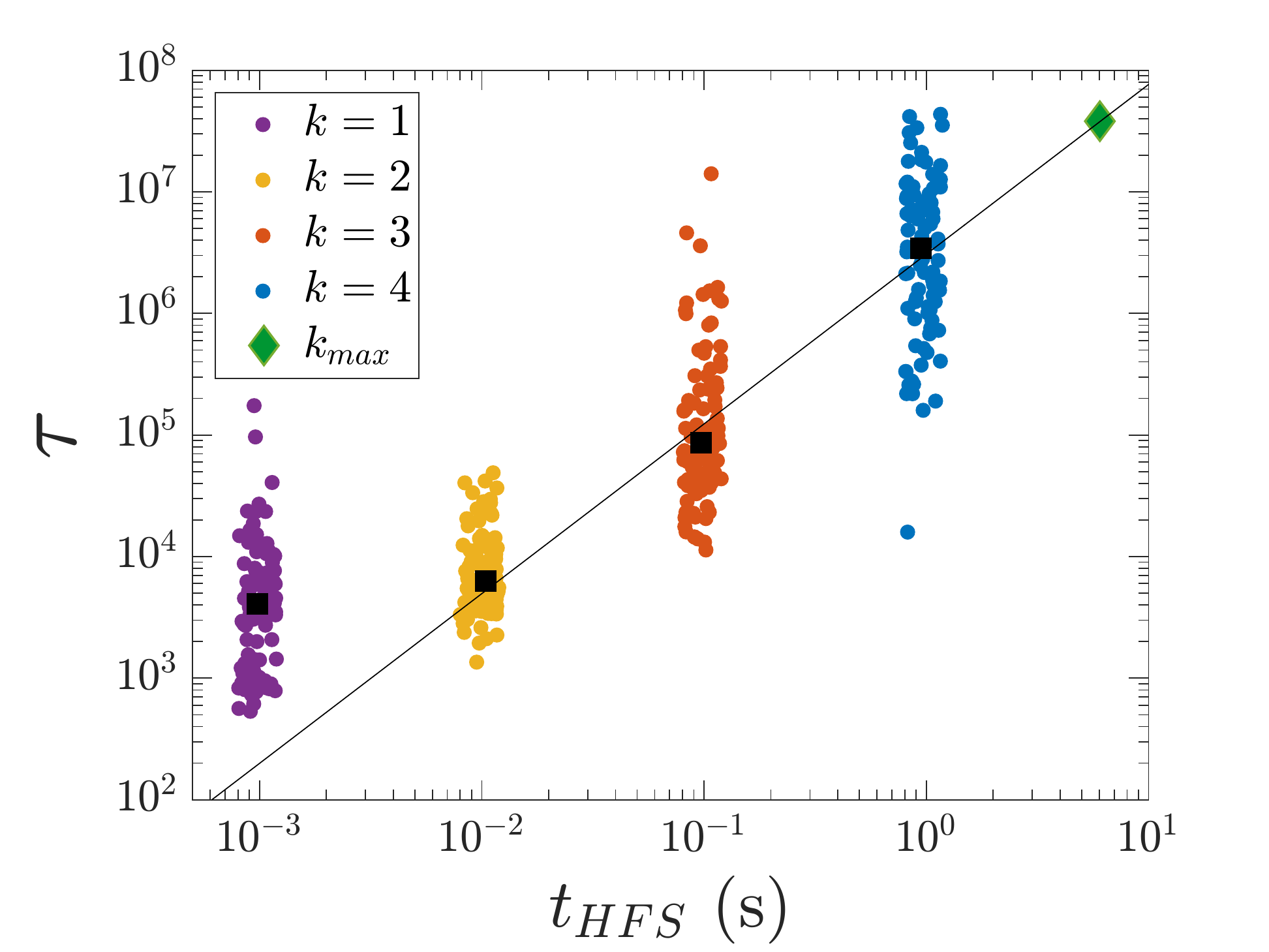}
\caption{(Color online) {\bf PT-hardness (mixing time) vs HFS-hardness.} The 400 instances generated as documented in the main text, examined on parallel tempering (PT). The linear fit of slope 1.40 is obtained from a least squares fitting on the median data point (black squares) of each data group, ignoring the $k=1$ group. We also include, indicated by a green diamond marker, the hardest instance ($k_{max}$) found using our adaptive algorithm, extrapolated using the linear fit to obtain the corresponding value for $\tau \approx 3.7\times10^7$. $\tau$ is given in units of Metropolis
sweeps.}
\label{s_vs_PT:graph}
\end{figure}

To establish the inherent hardness of the instances generated by the RAO algorithm, we have directly compared their time-to-solution $t_{HFS}$ to the PT mixing time, $\tau$. This is depicted in Fig.~\ref{s_vs_PT:graph}. Despite the apparent fluctuations, we observe an agreement between these two vastly different solvers, which can be quantified by the dependence $\tau \sim t_{HFS}^{1.4}$ (as measured by the median data point for the $k=2,3,4$ groups). This correlation has in fact allowed us to generate 14 
instances (out of the 100 of the $k=4$ group) with $\tau > 5 \times 10^7$ EMCS, which using straightforward `mining' would have required the generation and subsequent analysis of more than $2\times
10^6$ randomly generated instances (as found by Ref.~\cite{scirep15:Martin-Mayor_Hen}) --- about 100 times more costly in terms of computational resources~\cite{RAO_generation,PT_details}. Comparing this to the numerics quoted in the previous subsection (re. generating $\tau>10^7$ instances), we see RAO becomes even more beneficial over conventional methods as the problem difficulty bar is raised, and is very likely to be the \emph{only} way to obtain a large number of such temperature-chaotic instances.

 \par
\begin{figure}[h]
\centering
\includegraphics[scale=0.47]{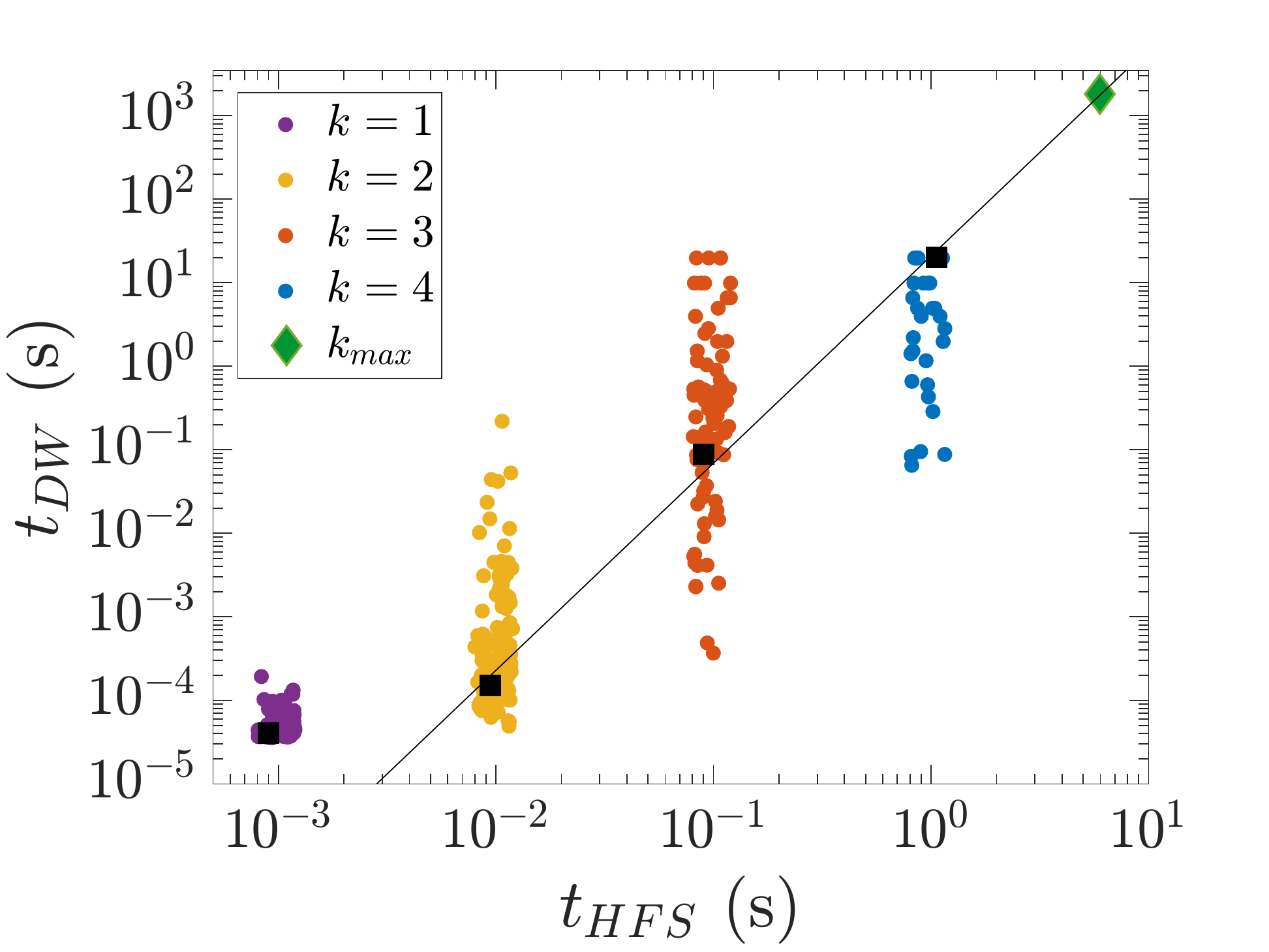}
\caption{(Color online) {\bf DW2-hardness vs HFS-hardness.} We plot the D-Wave TTS, $t_{DW}$, against $t_{HFS}$ for the 400 problem instances as explained in the main text. Note that there are fewer blue data points ($k=4$), compared to the others. This is because many of these instances were not solved once by the D-Wave machine in 660,000 attempts and so are left off of this graph. The linear fit of slope 2.48 is obtained from a least squares fitting on the lower quartile data point (black squares) of each data group, excluding the $k=1$ group. We also include, indicated by a green diamond marker, the hardest instance ($k_{max}$) found using our adaptive algorithm, extrapolated using the linear fit to obtain the corresponding value for $t_{DW}\approx10^3 s$, i.e., require $\sim 10^8$ attempts.}
\label{d_vs_s:graph}
\end{figure}

We perform a similar analysis using the D-Wave Two QA optimizer as the comparison platform. We define TTS as measured by DW2, $t_{DW}$, as the anneal time divided by the probability of successfully finding the GS. To establish probabilities of success, we ran each of the 400 instances on the D-Wave processor for roughly 650,000 anneals with individual anneal times in the range 20-40$\mu s$. For the hardest instances according to HFS ($k=4$), about 75\% of the instances were not solved even once by DW2. For this reason we use the lower quartile as a representative data point in Fig~\ref{d_vs_s:graph}.
\par
Here too, as with the PT comparison, large variations in the data are observed. Nonetheless, we see a strong correlation between the two solvers on average (as we expect), with $t_{DW} \sim t_{HFS}^{2.48}$. That DW2 scales unfavorably with HFS-hardness as compared to how the scaling of PT mixing time suggests that the QA chip may be detrimentally affected by `classical causes' such as thermal hardness of instances. 
\par
The hardest HFS instance found, $k_{max}$, neatly demonstrates the capabilities of the RAO algorithm applied to this particular graph. Instances of this type we estimate to have $\tau \approx 3.7 \times 10^7$, and $t_{DW} \approx 10^3 \,s$ (that is, would require $\sim 10^8$ DW2 anneals).
\par
As mentioned above, the HFS `easiest' instances ($k=1$) do not seem correlate with the other data groups (as we also see in Figs.~\ref{overlap:graph} and \ref{s_vs_PT:graph}). In fact the reason here is trivial; since $t_{DW}$ has a minimum equal to the median annealing time, $30 \mu s \approx 3\times10^{-5} s$ in this work, the easiest instances accumulate at this value, as is seen in Fig.~\ref{d_vs_s:graph}. We believe there may be an equivalent scenario occurring for PT, i.e., practical lower bound on $\tau$ (though, note, quantifying such a bound is non-trivial).

\par
These above correlation analyses all suggest that instances generated using our
RAO algorithm are indeed inherently more difficult optimization problems,
compared with randomly generated instances. In particular, RAO problems are
more akin to rare events (i.e., the instances displaying the strongest
temperature chaos in a large set of randomly chosen problems). For example,
the hardest RAO instances, indexed here by $k=4$ (equivalently,
$\tau_{HFS}\approx 10^0\, s$), are in general harder for the D-Wave Two
quantum annealer, as well as for parallel tempering algorithms. Moreover, we
see that (Fig.~\ref{overlap:graph}) these instances are thermodynamically more
difficult, with larger Hamming distances between the ground and dominant
first-excited states.

\section{Hard benchmarks with known ground state configurations}
\label{sect:planted}

The generation of instances with random couplings does not allow us in general to know the GS energy of the instances with certainty -- an important feature when carrying out comparison tests. Therefore, this section is devoted to another adaptive technique, building on work from Ref.~\cite{hen:15}, which generates hard instances, but also crucially allows for knowledge of the GS energy, without having to resort to exact solvers.

We apply our method to Ising-type instances with \emph{planted solutions}--- an idea borrowed from constraint satisfaction (SAT) problems~\cite{hen:15,Barthel:2002tw,Krzakala:2009qo}. 
Instances of this type are constructed around some arbitrary solution, by splitting the full graph up into smaller subgraphs, i.e., the Hamiltonian is written as a sum of small subgraph Ising Hamiltonians, $H = \sum_{j=1}^M H_j$. 
The coupling values of each sub-Ising Hamiltonian are chosen so that the planted solution is a simultaneous GS of all of the $H_j$, and therefore is also a GS of the total Hamiltonian $H$.
This knowledge circumvents the need for exact (provable) solvers, which rapidly become too expensive computationally as the number of variables grows, and as such is very suitable for benchmarking. 
In what follows, we shall choose our subgraph Hamiltonians to be randomly placed \emph{frustrated} cycles, or loops, along the edges of the hardware graph~\cite{hen:15} such that no configuration of the variables simultaneously minimizes all terms in the cost function (see Fig. 1 of \cite{hen:15} for examples of Hamiltonian loops on the DW2 graph). This frustration is known to often cause classical algorithms to get stuck in local minima, since the global minimum of the problem satisfies only a fraction of the Ising couplings and/or local fields~\cite{Binder86,Fischer}.

Unlike the signed $J_{ij}=\pm1$ problems studied above, planted-solution problems allow for the computation of certain measures of frustration (the reader is referred to Ref.~\cite{hen:15} for further details, and results pertaining to this approach in the context of benchmarking of experimental quantum annealers).
By combining the planted solution technique with the RAO method, one can generate instances which are harder than is possible to generate by simply using randomly placed loops on the graph, with the added benefit of still knowing the GS energy.
\par
We initialize the setup as in the above, i.e., we first pick a random planted solution, and place $M$ random sub-Hamiltonian loops (either frustrated or not) on the graph which satisfy the solution. This method allows us to easily calculate the GS energy as a sum of the individual loop energies with respect to the planted solution.
At variance with the RAO method, update attempts now instead of involving single random edges, involve Hamiltonian loops. We remove a random loop from the instance and add a new random loop, making sure to keep track of the GS energy, and making sure the new loop respects the planted solution.
\begin{algorithm}[H]
\caption{Loop Adaptive Optimization (LAO)}
\label{LAO}
\begin{algorithmic}[1]
\Procedure{GenerateFrustratedProblem}{}
\State Generate (random) solution
\State Place M random loops on graph, each respecting the planted solution
\State Calculate GS energy
\For{step = 1 \textbf{to} NSTEP }
\State Remove random loop from current instance
\State Pick new random loop and add, respecting planted solution
\State Get new TTS
\If{TTS increases}
\State Accept Change, update GS energy
\Else
\State Accept with probability $e^{-\beta |\Delta \text{TTS}|}$
\State Update GS energy if accepted
\EndIf
\EndFor
\EndProcedure
\end{algorithmic}
\end{algorithm}

One now has many different parameters affecting the performance, e.g., the total number of loops in the instance, the ratio of different sized loops (e.g., one can use a mix of size 4 and 6 loops etc.), different (positive) weights on the loops. One can also scrutinize the position of each loop to try to maximize e.g., the frustration (note that randomly adding loops can have the affect so as to cancel out frustration). Also, of course, similar comments about adjusting the algorithm as mentioned in the RAO section still apply here.

\begin{figure}[h]
\includegraphics[scale=0.47]{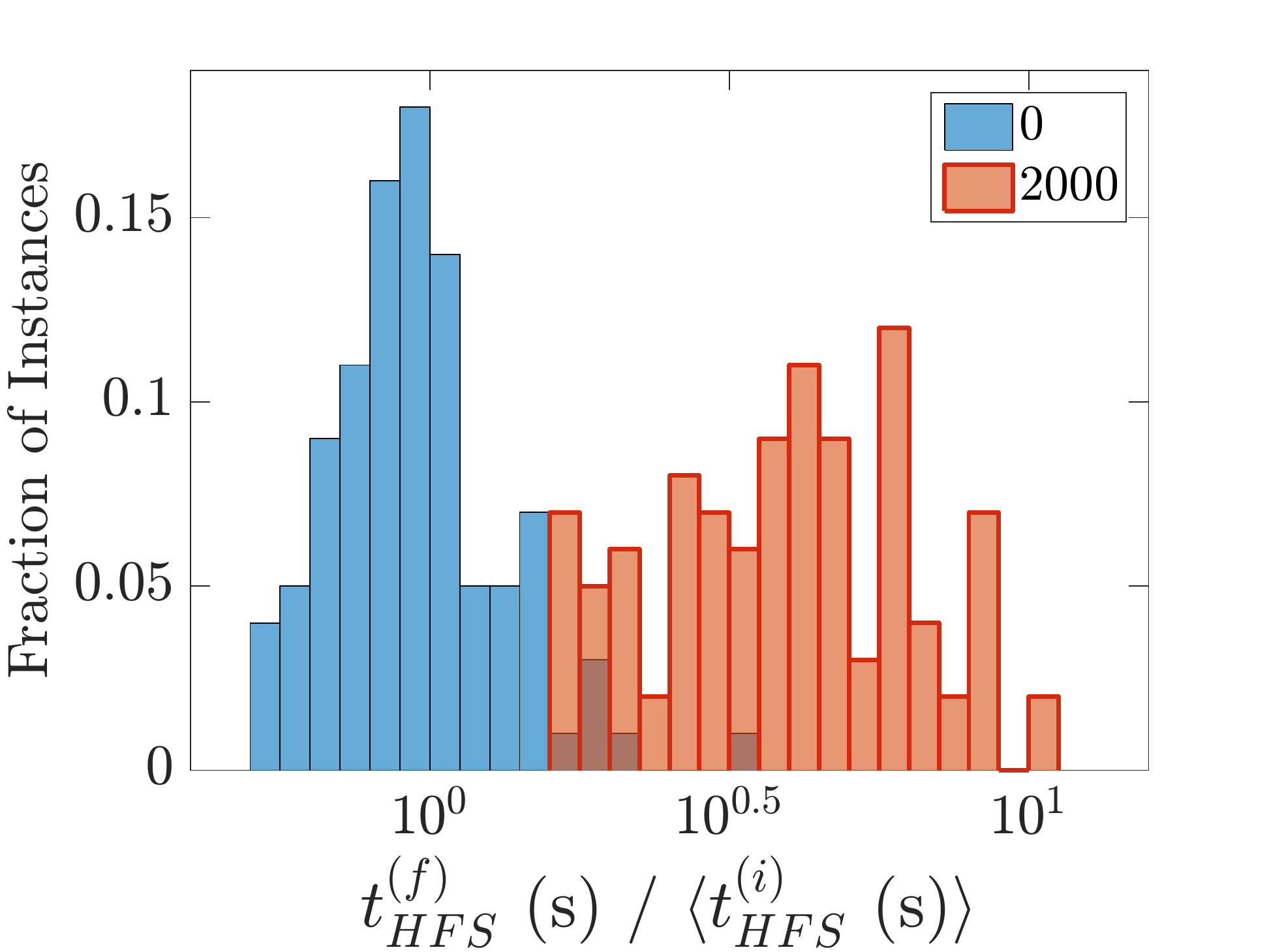}
\caption{(Color online) {\bf Histogram of the ratio of final to mean initial $t_{HFS}$ for 100 planted-solution instances (504-bit Chimera graph)}. We compare the LAO algorithm after 2000 steps (red) to the 100 random initial (blue) planted-solution instances, each containing 350 random loops. Updates consisted of adding and removing random loops, where size 4 (6) loops have a probability of 0.1 (0.9) being chosen, with integer loop weight chosen randomly in range [1,5]. We plot the histogram of the ratio final HFS TTS, $t^{(f)}_{HFS}$, after 0 (blue) and 2000 (red) LAO steps, to the mean initial TTS, $\langle t^{(i)}_{HFS} \rangle$. The mean TTS ratio after the 2000 steps is $\approx 4.3$, and the maximum final TTS is $\approx 0.15\,s$.}
\label{LAO_ratio:graph}
\end{figure}

\par
In Fig.~\ref{LAO_ratio:graph} we perform one such version of our LAO algorithm on 100 instances, and compare the final TTS to the typical TTS a random instance. While there is a general increase in problem difficulty, from that of a random instance, it is by far less than the equivalent figure for random signed instances (see Fig.~\ref{histogram:graph}).  
Note however that planted-solution problems may be tuned in numerous ways (see previous paragraph) to provide varying degrees of hardness, thereby altering the structure of the problem (in fact, they may be tuned to be harder than random signed instances~\cite{hen:15}). Thus, the choice of parameters, which we have not optimized here, may heavily affect the performance of LAO. Nevertheless, we have demonstrated the successful application of our main algorithm to instances with planted solutions, allowing for the generation of instances that are about an order of magnitude more difficult.

\section{Summary and conclusions}
\label{sect:conc}
We developed a technique to practically engineer extremely hard optimization problems to address the challenge of generating proper benchmarks for the testing of experimental quantum annealers. This was accomplished by treating the generation of hard problem instances as an optimization problem and subsequently devising a heuristic optimization algorithm to solve it. We demonstrated that one can successfully engineer Ising-type optimization problems with varying degrees of difficulty, defined by some suitable choice of  problem hardness.  To establish and confirm the inherent hardness of the instances, we measured the correlation between various independent measures of problem difficulty, in particular, from parallel tempering configurations we computed a Hamming distance measure between the ground and main first-excited states.
\par
We illustrated the ability to generate signed (i.e., $J_{ij}=\pm 1$), $512$-bit instances for the D-Wave Two Chimera which are greater in difficulty (as measured by the TTS of a very successful classical HFS solver) by more than two orders of magnitude compared to randomly generated instances. As designed, these instances were found to be more difficult both for the DW2 processor and for Parallel Tempering algorithms to solve. We have further shown that our technique is significantly faster than straightforward mining for hard instances which requires the generation and subsequent costly analysis of very many random instances.
\par
Since in designing benchmark tests it is often desirable, and necessary, to know with certainty the GS energy of the instances used, we have devised an adaptive technique which also allows one to generate hard instances for which a GS is known, based on problems with planted solutions~\cite{hen:15}. While in this case the method is somewhat less effective, it nonetheless allows one to easily generate problem instances that are rare to find by random generation of instances. 
\par
The generation of hard instances is one of the key tools to understand some fundamental, but unanswered questions in spin glass theory as well as in the field of quantum annealing. For example, what makes certain problems hard? What are the most reliable ways to classify problem difficulty? Is there a marked difference between quantum and classical hardness? 
The techniques presented in this work may be further utilized to the end of observing the elusive quantum speedup (provided that there could be one). The adaptive generation of hard instances may be further leveraged to systematically study the properties (geometric, thermodynamical or otherwise) of the resultant instances, paving the way towards the systematic generation of inherently hard spin-glass instances. 
\par
By simultaneously updating an instance, one may further try to use as a figure of merit that is to be maximized, the ratio of the `classical TTS' to quantum (e.g., experimental annealer) TTS. This may hopefully allow for the engineering of instances which are classically-hard but quantum-easy. These may consequently be studied so as to enhance our understanding of the differences between quantum and classical hardness. 

\section{Acknowledgements}
We thank Ehsan Khatami for useful comments and suggestions.

This work was partially supported by MINECO (Spain) through Grant Nos. FIS2012-35719-C02,  
FIS2015-65078-C2-1-P. Computation for the work described in this paper was partially supported by the University of Southern California’s Center for High-Performance Computing (http://hpcc.usc.edu).

\bibliography{refs}

\begin{thebibliography}{59}
\bibitem [{RAO()}]{RAO_generation}%
  \BibitemOpen
  \href@noop {} {}\bibinfo {note} {{To generate the hardest instances, $k=4$,
  we ran our algorithm 780 times with up to 1000 steps (that is, at least 100
  of the 780 instances fell within the time interval [0.8,1.2]$s$). The total
  CPU time this took was approximately 400 hours (though note that our code was
  not optimized). Also note that of these 780, about 120 have $t_{HFS}>1.2s$.
  The other $k$ groups took significantly less time to generate.}}\BibitemShut
  {Stop}%
\bibitem [{PT_()}]{PT_details}%
  \BibitemOpen
  \href@noop {} {}\bibinfo {note} {{Screening instances on PT occurs in 2
  phases: Phase 1 screens 76800 random instances, taking $\sim 900$ CPU hours.
  Phase 2 then screens the 1024 (estimated) worst case instances for another
  100 hours. Given 76800 random 500+ bit instances we expect approximately 15
  to have $\tau>10^7$ ($\sim 65$ hours per instance).}}\BibitemShut {Stop}%
\end{thebibliography}
\pagebreak

\appendix

\section{\label{app:dwave} D-WAVE TWO ANNEALER}

\begin{figure}[hbp]
\includegraphics[ scale=0.4 ]{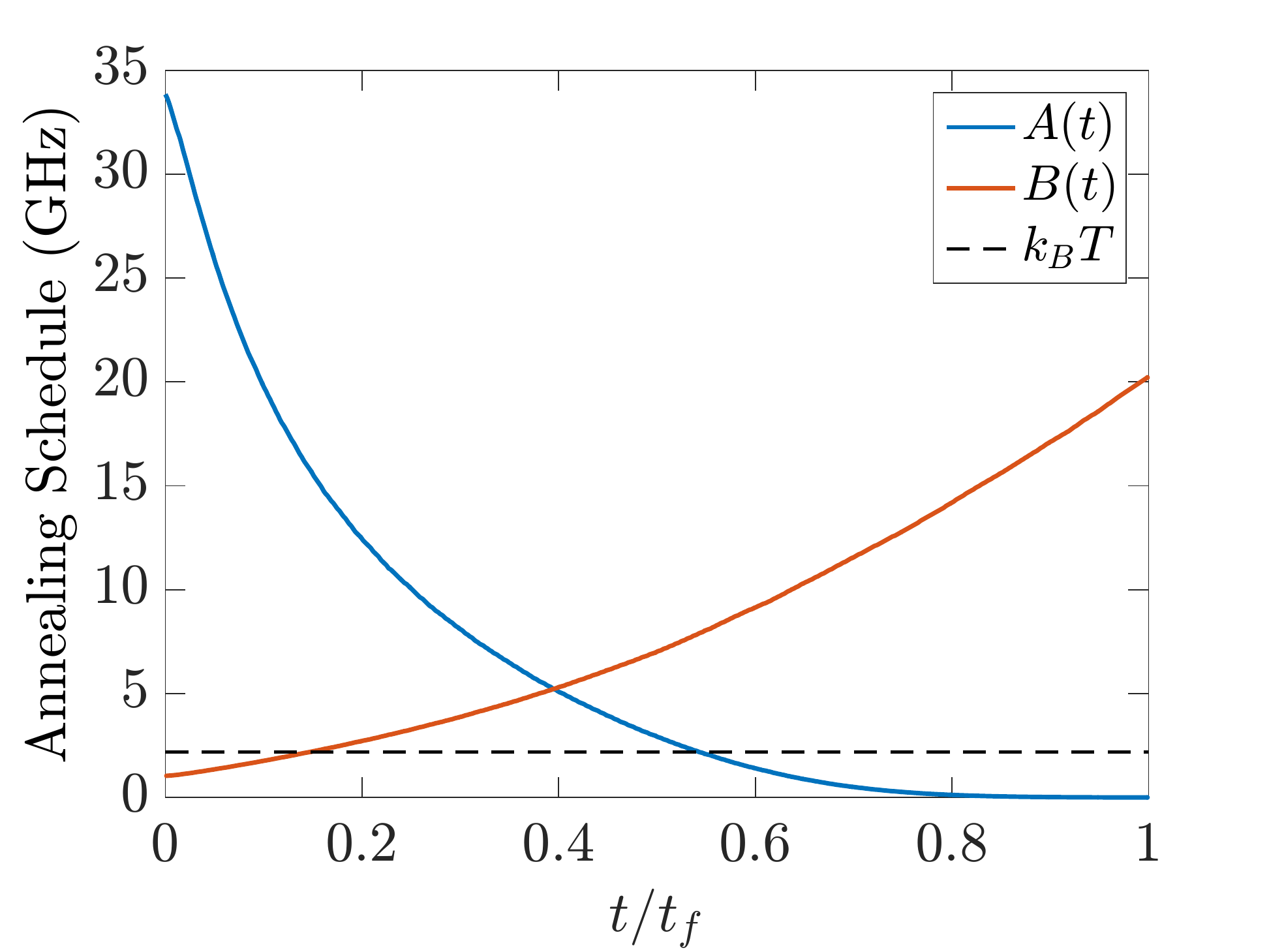}
\caption{(Color online) \textbf{Annealing schedule of the DW2 device.}  The annealing curves $A(t)$ and $B(t)$ are calculated using rf-SQUID models with independently calibrated qubit parameters. Units of $\hbar = 1$.  The operating temperature of $17$mK is also shown.}
\label{fig:schedule}
\end{figure}

\begin{figure}[htp]
\includegraphics[angle=0,width=0.7\columnwidth ]{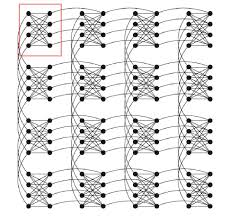}
\caption{(Color online) {\bf A 4 by 4 patch of the full 8 by 8 Chimera graph for the DW2 chip}. Top left (red) shows a single $K_{4,4}$ bipartite graph. The qubits or spin variables occupy the vertices (circles) and the couplings $J_{ij}$ are along the edges. 
Of the $512$ qubits, $504$ were operative in our experiments. }
\label{fig:chimera}
\end{figure}

The D-Wave Two (DW2) is marketed by D-Wave Systems Inc. as a quantum annealer, which evolves a physical system of superconducting flux qubits according to the time-dependent Hamiltonian
\be \label{eq:Hquantum}
H(t) = A(t) \sum_{i\in V} \sigma_i^x + B(t) H_p \ , \quad t\in[0,t_f] \ ,
\ee
with $H_p$ given in Eq.~\eqref{Ising}. The annealing schedules given by $A(t)$ and $B(t)$ are shown in Fig.~\ref{fig:schedule}.  Our experiments used the DW2 device housed at the USC Information Sciences Institute, with an operating temperature $\approx17$mK. 
The Chimera graph of the DW2 used in our work is shown in Figure~\ref{fig:chimera}. Each unit cell is a balanced $K_{4,4}$ bipartite graph. In the ideal Chimera graph (of $512$ qubits) the degree of each vertex is $6$ (except for the corner unit cells). In the actual DW2 device we used, $504$ qubits were functional.

\section{IMPLEMENTATION DETAILS}
\label{app:implementation}
All of our numerical results were obtained using A. Selby's (heuristic) version of HFS~\cite{selby:13a,Selby:2014tx} (i.e., with input settings -S3 -m1), running on a single core of an Intel Xeon CPU E5-1650 v2 running at 3.50GHz. With these settings, the code will halt only after it has found agreement in the lowest energy for Eq. \ref{Ising}, $n+1$ consecutive times (option -p  set to $n$ in Selby's code), over independent HFS sweeps (Selby's code technically solves Quadratic Unconstrained Binary Optimization (QUBO) problems, but the mapping from Ising problems of this sort is trivial).
\par
We define the HFS time-to-solution, $t_{HFS}$ in the main text, as $t_{HFS}:=\lim_{n \rightarrow \infty}T_{\text{step}}/(n+1)$, where $T_{\text{step}}$, depending on $n$, is the physical wallclock run time of Selby's code, for a single QUBO instance (also see footnote~\cite{tts_hfs}). That is, $T_{\text{step}}$ is the time taken to find agreement in the lowest energy $n+1$ times in a row, and as such we define this quantity as $T_{\text{step}}:=\sum_{i=0}^n t_i$, where $t_i$ is the time taken to find $i$th ($i>0$) occurrence of the (presumed) minima, and $t_0$ the time to first detect this minima. In practice, to obtain a reasonable estimate of $t_{HFS}$ one should take a `large' value for $n$, e.g., $n>500$.

\par
We face two practical issues: 1) Fixing some value for $n$, and adapting the instance under RAO, of course means that the wallclock runtime of each step $T_{\text{step}}$ of RAO increases as the problem becomes harder, meaning for a large number of RAO steps, the algorithm may take a very long time to complete. 2) We noticed that in addition to $t_{HFS}$ increasing, so do the differences, $|\Delta t_{HFS}|$, and as such, the acceptance probability, $e^{-\beta |\Delta t_{HFS}|}$, may quickly become negligible. We provide a quick and easy solution to these two problems, by varying just one parameter, $n$. This is by no means an optimal solution, but it has enabled us to generate many hard instances with fewer computational resources compared to what would otherwise be required. 

By defining a cutoff return time, $T_{\text{max}}$ (we picked $T_{\text{max}}=3s$), such that if $T_{\text{step}} > T_{\text{max}}$, we reduce $n\rightarrow n/2$  (note we never let $n$ be lower than 16, and initialize it with $n=512$). This allows us to control the growth of $T_{\text{step}}$, and hence bound (at least somewhat) the total runtime of the RAO algorithm. The accuracy in the estimation of $t_{HFS}$ of course decreases as $n$ decreases, therefore one should run the final instance (i.e., the instance after adapting it as per the RAO algorithm) with some large choice of $n$. 
\par
In addition, we let $\beta$ depend linearly on $n$, so that as $t_{HFS}$ increases (hence $|\Delta t_{HFS}|$ too), $\beta$ is lowered, and as such we can control the acceptance probability (again, at least somewhat). Our particular choice used to generate the 100 hardest instances ($k=4$) for the results section was $\beta=6.5 \cdot n$. The value 6.5 is not particularly special; our version of RAO seemed to perform best with this choice over a small trial of other values.


\end{document}